\documentclass[preprint]{aastex}

\usepackage{amssymb}
\usepackage{graphicx}

\title{Understanding the Behavior of Prometheus and Pandora}
\author{Alison J. Farmer, Peter Goldreich}
\affil{Theoretical Astrophysics, MC 130-33, Caltech,
  Pasadena, CA 91125}
\affil{Institute for Advanced Study, Einstein Drive,
  Princeton, NJ 08540}
\email{ajf,pmg@ias.edu}
\altaffiltext{0}{21 pages, 2 tables, 13 figures \\
\emph{Running head:} Understanding Prometheus and Pandora\\
\emph{Corresponding author:} Alison Farmer, MS-51, Center for Astrophysics, 60 Garden St., Cambridge, MA 02138, afarmer@cfa.harvard.edu}

\begin{document}

\begin{abstract}

We revisit the dynamics of Prometheus and Pandora, two small moons
flanking Saturn's F ring. Departures of their orbits from freely
precessing ellipses result from mutual interactions via their 121:118
mean motion resonance. Motions are chaotic because the resonance is
split into four overlapping components. Orbital longitudes were
observed to drift away from predictions based on \emph{Voyager}
ephemerides. A sudden jump in mean motions took place close to the
time at which the orbits' apses were antialigned in 2000. Numerical
integrations reproduce both the longitude drifts and the jumps. The
latter have been attributed to the greater strength of interactions
near apse antialignment (every 6.2 years), and it has been assumed
that this drift-jump behavior will continue indefinitely.

We re-examine the dynamics of the Prometheus-Pandora system by analogy
with that of a nearly adiabatic, parametric pendulum. In terms of this
analogy, the current value of the action of the satellite system is
close to its maximum in the chaotic zone. Consequently, at
present, the two separatrix crossings per precessional cycle occur
close to apse antialignment. In this state libration only occurs when
the potential's amplitude is nearly maximal, and the ``jumps'' in mean
motion arise during the short intervals of libration that separate
long stretches of circulation. Because chaotic systems explore the
entire region of phase space available to them, we expect that at
other times the Prometheus-Pandora system would be found in states of
medium or low action.  In a low action state it would spend most of
the time in libration, and separatrix crossings would occur near
apse \emph{alignment}. We predict that transitions between these different
states can happen in as little as a decade. Therefore, it is incorrect
to assume that sudden changes in the orbits only happen near apse
antialignment.

\end{abstract}

\newpage \vspace{10in}
\section{Introduction}

Discovered by the \emph{Voyager} spacecraft in 1980 and 1981,
Prometheus and Pandora are two small moons of Saturn. Properties of
their orbits are summarized in Table \ref{tab:eqterms}. A series of
observations starting in 1995 found that the orbital longitude of each
satellite deviated by about 20 degrees from its value as predicted by
the \emph{Voyager} ephemeris (Bosh \& Rivkin, 1996; Nicholson et al.,
1996; McGhee et al., 2001). Further, each satellite's mean motion
underwent an abrupt change in 2000 (French et al., 2002), seen as
``kinks'' in their mean longitudes. Goldreich \&
Rappaport (2003a; GR03a) pinpointed chaos due to the satellites' mutual
gravitational interactions as the cause of these discrepancies. The
system was found to have a Lyapunov exponent $\sim$0.3 yr$^{-1}$. Jumps in
the mean motion were attributed to the stronger interactions that
occur when the orbits' apses are antialigned. Orbital integrations in
GR03a reproduce the observed gradual drifts away from the
\emph{Voyager}-based predictions. Kinks in the mean longitudes are
apparent every 6.2 years, at times of apse antialignment.  Goldreich
\& Rappaport (2003b; GR03b) captures the essential dynamics of the
system by including only interactions due to the 121:118 mean motion
resonance.  More complete numerical simulations of the system, which
include the influences of other Saturnian satellites, have been
performed by Cooper \& Murray (2004); Jacobson \& French (2004); and
Renner, Sicardy \& French (2005). These confirm that the simplified
dynamics in GR03b is sufficient to describe the chaotic motions of
Prometheus and Pandora.


Physical explanations for the mean motion jumps have not advanced
beyond those proposed by Goldreich \& Rappaport: all of the above
papers cite enhanced interactions at apse antialignment as the reason
for the kinks in the mean longitudes observed in 2000. They do nothing to
dispel the expectation that the mean longitudes will continue to
display the same drift-kink behavior indefinitely into the future.  A
practical consequence of this belief is that orbits fitted to Cassini
data avoid times around apse antialignment and assume freely
precessing ellipses between these times (Porco et al. 2005).

We show that a proper understanding of the dynamics of the
Prometheus-Pandora system is more subtle than previously
recognized. In doing so we exploit an analogy with the dynamics of a
parametric pendulum to reinterpret numerical integrations presented in
GR03b. We confirm that separatrix crossings are the cause of chaos and
that their rate determines the magnitude of the Lyapunov exponent.
Our focus is on the precessional phase at which separatrix crossings
occur. Currently the two that take place during each 6.2 yr
precessional cycle occur near apse antialignment and the kinks arise
during intervals of libration separated by long stretches of
circulation. More generally, separatrix crossings can occur at any
precessional phase, with large changes in these phases predicted on
timescales of one to two decades. When crossings take place near apse
alignment, the system remains chaotic but displays qualitatively
different features than at present. 

This paper is arranged as follows. In \S \ref{sec:eq} we
demonstrate the similarity between the dynamics of the
Prometheus-Pandora interaction and that of a parametric pendulum. We
study the behavior of an adiabatic parametric pendulum in
\S \ref{sec:behave}. In \S \ref{sec:panprom} we apply these findings to
the Prometheus-Pandora system, and discuss both the implications and
the limitations of our analogy. We conclude in \S \ref{sec:conc}.

\section{Similarity of equations}
\label{sec:eq}

GR03b showed that the chaotic motions of Prometheus and Pandora arise
from their 121:118 mean motion resonance. Differential precession of
the two eccentric orbits splits the resonance into four discrete
components whose arguments are $\psi_q=\psi-\delta_q$, with
\begin{equation}
\psi\equiv 121\lambda-118\lambda',
\label{eq:psi}
\end{equation}
where $\lambda$ and $\lambda'$ are
respectively the mean longitudes of Prometheus and Pandora, and
\begin{equation}
\delta_1\equiv 3\varpi, \;
\delta_2\equiv 2\varpi+\varpi',\;\delta_3\equiv \varpi+2 \varpi',\;
\delta_4\equiv 3\varpi'.
\end{equation}
GR03b note that because of the rapid precession caused by
Saturn's oblateness, interactions between the satellites produce
negligible effects on their apsidal angles and orbital
eccentricities. Therefore it is adequate to only treat changes in
mean motions, or equivalently in the angle $\psi$. 

The equation of motion for $\psi$ reads (GR03b)
\begin{equation}
\frac{d^2 \psi}{d t^2} = 3(121n')^2\frac{m}{M}\left[1+\frac{a m'}{a' m}\right]
\sum_{q=1}^4 C_q \sin(\psi-\delta_q),
\label{eq:pp_eom}
\end{equation}
where $M=6 \times 10^{29}$ g is the mass of Saturn, and where the
values of terms in Eq. \ref{eq:pp_eom} are given in Tables
\ref{tab:eqterms} and \ref{tab:res}.

\begin{figure}
\begin{center}
\includegraphics[width=0.5 \textwidth]{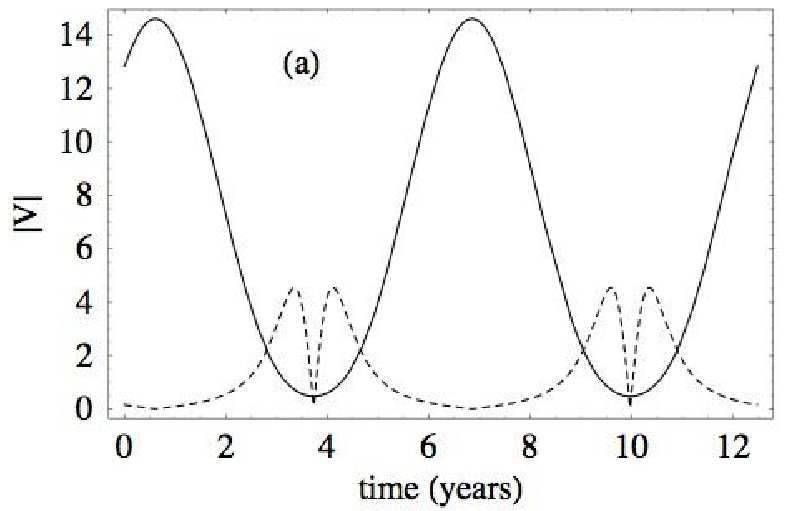}
\end{center}
\includegraphics[width=0.5 \textwidth]{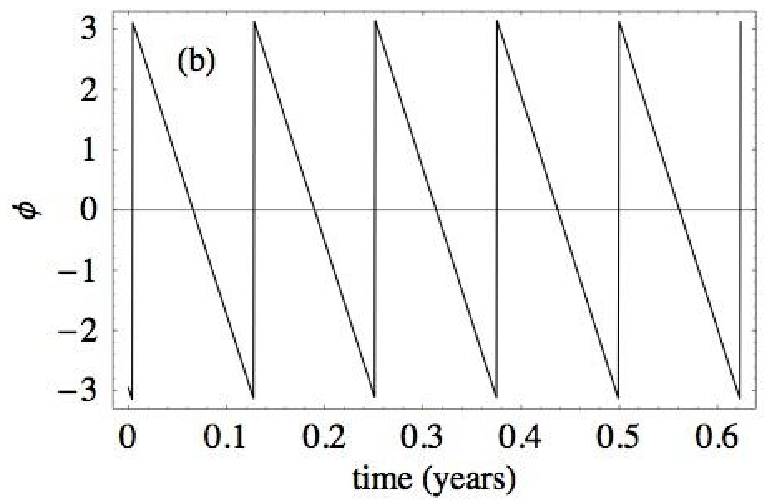}
\includegraphics[width=0.5 \textwidth]{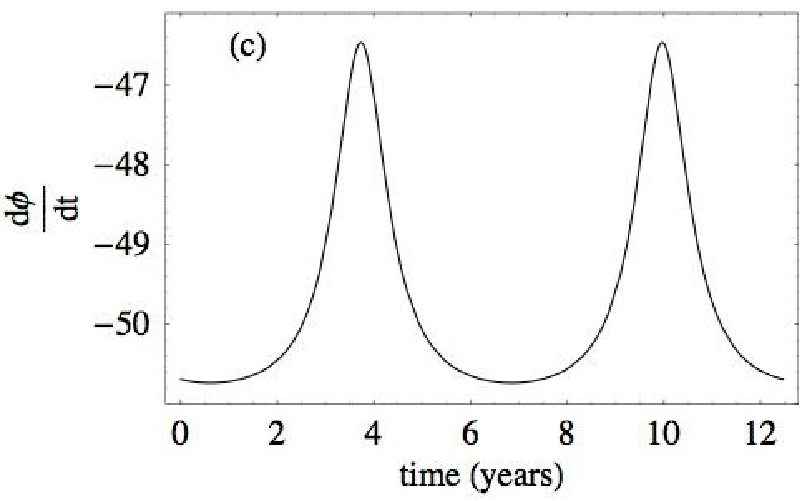}
\caption{Properties
of the potential in which the Prometheus-Pandora system moves: (a)
solid line--the amplitude of the potential as a function of time,
dashed line--$\ddot \phi$; (b) Drift phase $\phi$ as a function of
time; (c) Rate of change of the drift phase. This is the drift rate of
the hills in the potential. In most of this study we ignore the
variation of $\dot \phi$, since it is localized to a small fraction of
the period.}
\label{fig:pot}
\end{figure}

 Equation \ref{eq:pp_eom} can be written as
\begin{equation}
\frac{d^2 \psi}{d t^2} = -\omega_0^2 A(t) \sin[\psi-\phi(t)],
\end{equation}
at time $t$, where $\omega_0 \simeq 3.8 \rm{~yr}^{-1}$ and
$|A| \leq 1$.
Defining $\Psi=\psi-\phi$ and thus transforming to the frame drifting with the potential 
at $\dot \phi$,
we obtain
\begin{equation}
\frac{d^2 \Psi}{d t^2} = -\omega_0^2 A(t) \sin\Psi + \ddot{\phi}.
\label{eq:full}
\end{equation}
Provided $\ddot{\phi} \ll \omega_0^2 A$, we can drop the last term,
leaving
\begin{equation}
\frac{d^2 \Psi}{d t^2} = -\omega_0^2 A(t) \sin\Psi,
\label{eq:pp_approx}
\end{equation}
precisely the equation of motion of a parametric pendulum.
The potential in which $\Psi$ moves is given by
\begin{equation}
V(\Psi,t)=-\omega_0^2 A(t) \cos\Psi.
\end{equation} 
Plots of $|V|$, $\phi$, $\dot\phi$ and $\ddot{\phi}$ are displayed in
Figure \ref{fig:pot}.\footnote{Although $\ddot \phi \neq 0$, we
approximate $\dot \phi$ to be constant at $-50.7 \textrm{~yr}^{-1}$
when we calculate the action for the Prometheus-Pandora system. The pendulum analogy takes us a long way
toward understanding the behavior of the system. We assess its
validity in \S \ref{sec:probs}.}

\begin{figure}
\includegraphics[width=\textwidth]{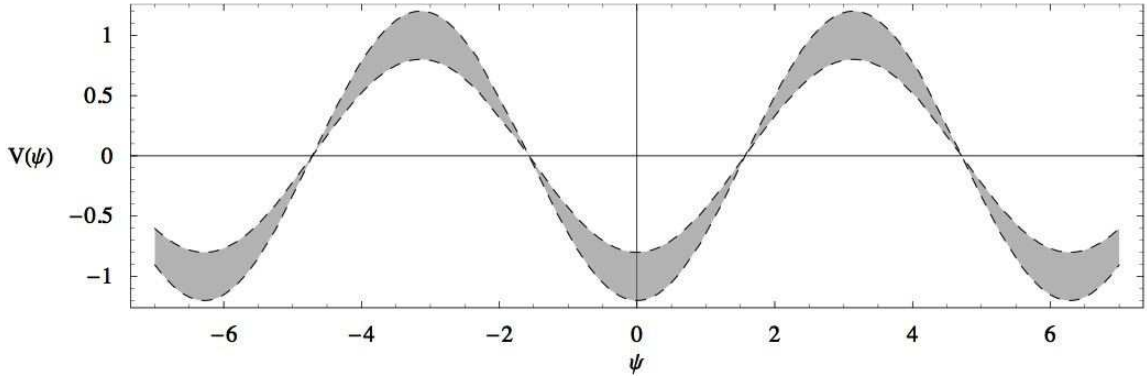}
\caption{The potential
  $V(\psi)$ in which the simplified parametric pendulum
  moves, for $\alpha=0.2$. Shaded area shows range over which
  potential varies with time.}
\label{fig:potrange}
\end{figure}

The timescale for variation of
$|V(t)|$ is $2\pi/|\dot \varpi-\dot\varpi'|\approx 6.2$ yrs, and the
fractional variation in its amplitude is of order unity. The typical
oscillation period of $\Psi$ is $\omega \sim 2\pi/\bar{|V|}^{1/2}\sim
2$ yrs, so the slowness parameter $\epsilon \sim 1\times 2/6.2 \simeq
0.3$.
Thus the Prometheus-Pandora system is only marginally
adiabatic. Rather than examine it directly, it proves instructive to
consider a parametric pendulum for which adiabaticity is robust. For this
we choose \begin{equation}A(t) = 1+\alpha \cos(\kappa t),\end{equation} where $\alpha < 1$ and
$\kappa \ll \omega_0$, so that for small oscillations the system is
adiabatic with slowness parameter \begin{equation}\epsilon \sim \alpha
\kappa/\omega_0 \ll 1.\end{equation} Without loss of generality we can set
$\omega_0 =1$. This potential is illustrated in Figure
\ref{fig:potrange}.

\subsubsection{Integration}
\label{sec:integrate}
All integrations in this paper are done in \emph{Mathematica 5} using
a Runge-Kutta routine
with fixed step size.
Care is taken to maintain numerical accuracy at a level such that
differences in initial conditions determine the rates at which
neighboring trajectories diverge.

\section{Behavior of the simplified parametric pendulum}
\label{sec:behave}

A simple pendulum can exhibit motion of two kinds:\footnote{for a
simple pendulum, $A$ is independent of time.}  libration for $E<|V|$
in which $\psi$ is trapped in a valley, and circulation for $E>|V|$ in
which $\psi$ passes over the hills. Between these two regimes lies the
separatrix, a singular trajectory of infinite period with energy 
\begin{equation}
E=\dot\psi^2/2-A\cos\psi=A.
\end{equation}

As $A(t)$ varies, the motion of a parametric pendulum may switch
between these two regimes. Transitions from large amplitude librations
to circulations may occur as the potential shallows. Likewise,
transitions in the opposite direction can take place as the potential
steepens. No matter how slowly the potential varies, adiabatic
invariance is violated during separatrix crossings.

\begin{figure}
\includegraphics[width=0.5 \textwidth]{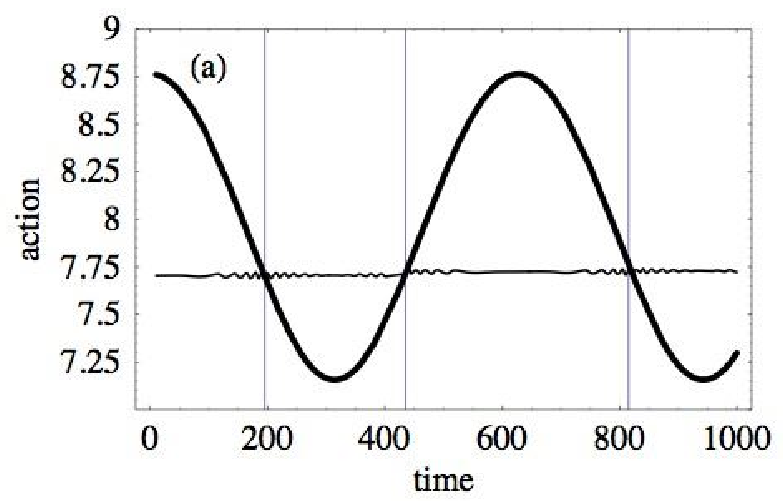}
\includegraphics[width=0.5 \textwidth]{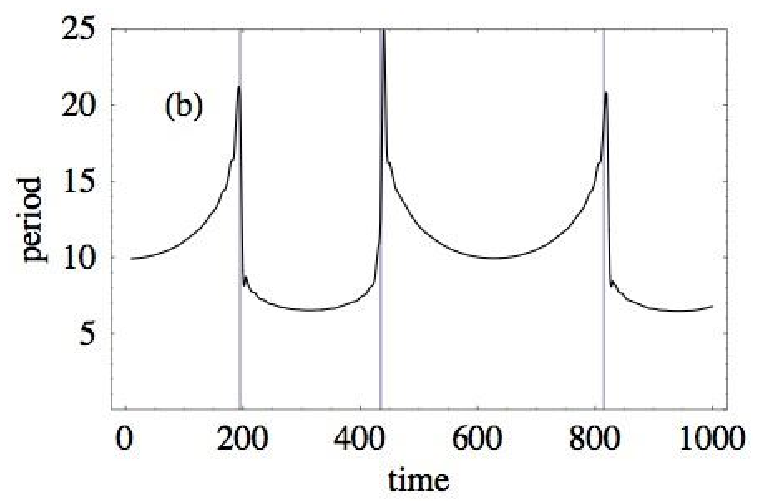}
\caption{For the model parametric pendulum system with  $\alpha=0.2$,
$\kappa=0.01$: (a) Action as a function of time; thick line is
separatrix action and thin line is system action. (b) System period as a
function of time, with vertical lines at locations of separatrix crossings, where the period becomes large.}
\label{fig:actions}
\end{figure}

\begin{figure}
\includegraphics[width=0.5 \textwidth]{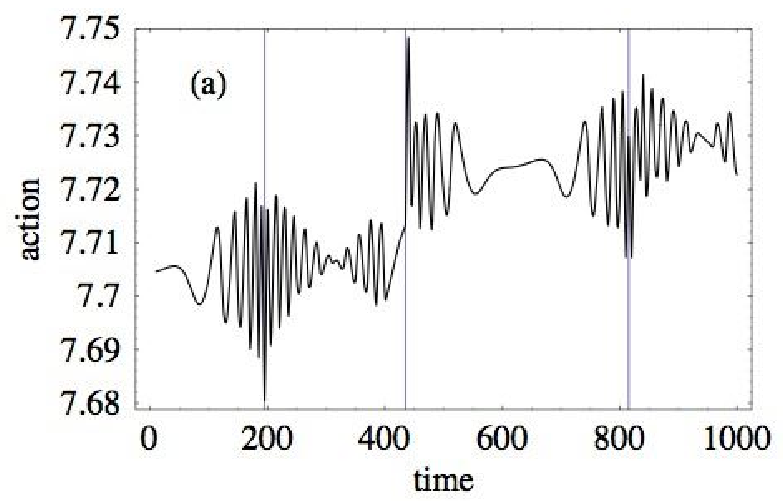}
\includegraphics[width=0.5 \textwidth]{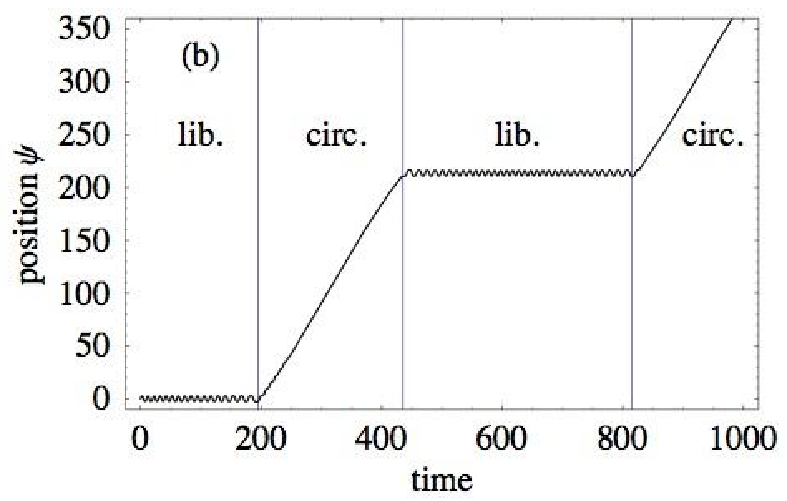}
\caption{For the same system as in Fig. \ref{fig:actions}:  (a) Zoom-in on system action,
showing jumps in action at separatrix crossings (marked by vertical lines).  (b) Evolution of
$\psi(t)$, showing transitions from libration to circulation and vice
versa at separatrix crossings.}
\label{fig:model_plc}
\end{figure}

For an adiabatic system the action, $J(t)$, remains nearly constant as
$A(t)$ varies.  We calculate $J(t)$ by freezing $A(t)$ and integrating
over a period;\footnote{When the system is in libration, we integrate
over only half a period of the motion, because otherwise the action
changes by a factor of 2 across the separatrix at fixed energy.}
\begin{equation}
J(t) = \oint p \; dq =\oint \dot \psi\;  d\psi = \oint \sqrt{2[E(t)-V(\psi')]} \; d\psi'.
\end{equation}
In a similar manner, we obtain the period from $P(t) = \oint dq/\dot q$.
On the separatrix,
\begin{equation}
J_{\rm sep}(t)=8 A(t)^{1/2}.
\label{eq:jsep}
\end{equation}
Figure \ref{fig:actions} illustrates some aspects of the behavior of a
parametric pendulum characterized by $\kappa=0.01$, $\alpha=0.2$,
i.e., slowness parameter $\epsilon \sim 0.002$.  We see that $J(t)$
maintains a nearly constant value as both the action on the separatrix
and the period undergo large variations. Jumps in action at separatrix
crossings are displayed at higher resolution in Figure
\ref{fig:model_plc} along with the evolution of $\psi(t)$.

\subsection{Separation in phase space and Lyapunov exponent}

\begin{figure}
\includegraphics[width=0.5 \textwidth]{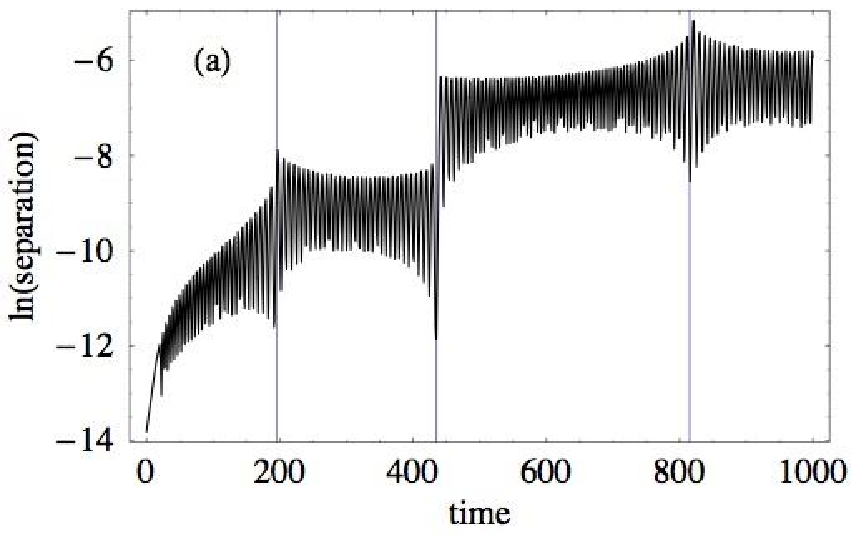}
\includegraphics[width=0.5 \textwidth]{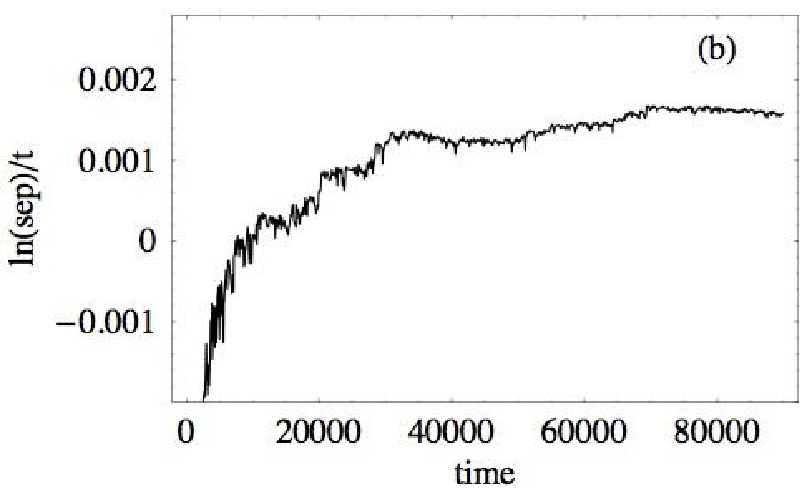}
\caption{For the same system as in Fig. \ref{fig:actions}, we illustrate the separation 
of neighboring trajectories in phase space. (a) A plot of the phase space separation as a function of time shows the jumps in both the separation and the rate of separation that occur at separatrix crossings.
 (b) A longer integration gives the system Lyapunov exponent $\ell$ as the asymptotic value of $\ln({\rm separation})/{\rm time}$ at late time, which from the plot is $\ell \sim 0.002$.}
\label{fig:model_sep}
\end{figure}

The rate at which neighboring trajectories in phase space separate is
of practical interest because it limits our ability to make
predictions about the future. For chaotic systems, small errors in
initial conditions amplify exponentially, invalidating long-term
predictions.  To track the separation of two neighboring trajectories,
we Taylor expand the equation of motion (Eq. \ref{eq:pp_approx}) to
first order in $\Delta \psi$:
\begin{equation}
\frac{d^2 \Delta \psi}{d t^2}=-A(t)\omega_0^2 \Delta \psi \cos\psi.
\label{eq:linear}
\end{equation}
The phase space separation,
\begin{equation}
S(t)=\left[\Delta\psi^2+\left(\frac{d\Delta\psi}{dt}\right)^2\right]^{1/2},
\end{equation}
is calculated by the simultaneous integration of
Eqs. \ref{eq:pp_approx} and \ref{eq:linear}. The Lyapunov exponent,
$\ell$, is the limit as $t\to \infty$ of $(\ln S)/t$. From Figure
\ref{fig:model_sep} we find $\ell \simeq 0.0022$, which is close to the
rate of separatrix crossings.

The Lyapunov exponent measures the average exponentiation rate of the
divergence of neighboring trajectories. Figure \ref{fig:model_sep}a
shows how this divergence proceeds through a few separatrix
crossings. At each crossing the separation $S$ between trajectories
undergoes a sudden jump, corresponding to the differential phase delay
introduced there between neighboring trajectories. Between crossings
$S(t)$ displays an oscillation superposed on a linear trend.  The
oscillation reflects the periodic motion of the pendulum, and the
linear variation is due to the period difference between neighboring
trajectories. Both the jumps in $S$ at separatrix crossings and the
linear variations between them produce an increase of $S$ on
average. Since on average $S$ and its slope change by a constant
fractional amount at separatrix crossings, $S(t)$ grows exponentially.

\subsection{Excursions in the chaotic zone}
\label{sec:excursions}
Only trajectories that come close to the separatrix are chaotic. The
range in action over which there is chaotic behavior is
\begin{equation}\Delta J_{\rm cz} \simeq 8(A_{\rm max}^{1/2}-A_{\rm
min}^{1/2}) \simeq 8 \alpha.\end{equation} A system that starts with
action outside this zone never crosses the separatrix and undergoes
regular motion. A system that starts within it eventually explores the
entire range of chaotic actions. Coverage of phase space is uniform in
the chaotic region.

Adiabaticity is violated along trajectories that cross a separatrix
and the action jumps by
\begin{equation}
\Delta J = -4\frac{\dot A}{A} \ln\left[2 \cos\left(\frac{\pi}{2}
\sin\frac{\psi_s}{2}\right)\right],
\label{eq:tim}
\end{equation}
where $\psi_s$ is the phase of the pendulum at which $E=E_{\rm sep}$
(Timofeev 1978). The validity of Eq. (\ref{eq:tim}) requires that
$\dot A$ be constant while the trajectory crosses the separatrix.
However, $\dot A$ varies on the timescale $\delta t=|\dot A/\ddot
A|=|\tan\kappa t/\kappa|$ which vanishes at the extrema of $A$. Thus
Eq. (\ref{eq:tim}) does not apply within boundary layers of width
$\delta J$ at the top and bottom of the chaotic zone. We estimate
$\delta J\approx C\alpha\kappa^2$, where $C$ is a (large)
dimensionless number.

\begin{figure}
\includegraphics[width=\textwidth]{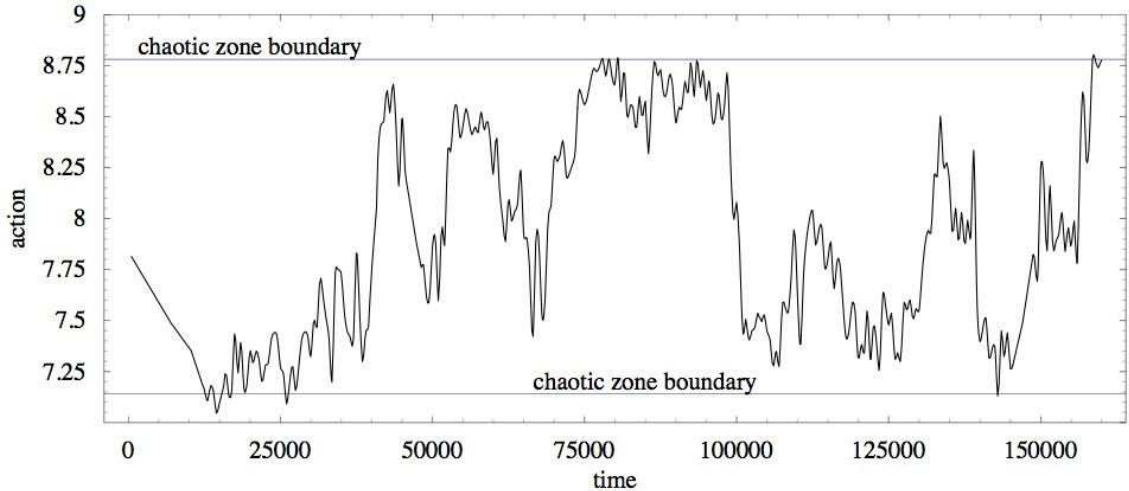}
\caption{Using a less adiabatic model parametric pendulum than in Fig. \ref{fig:actions} ($\kappa =0.1$, $\alpha=0.2$), we plot the value of the system action over about 5000 separatrix crossings. The action diffuses more rapidly than that in Fig. \ref{fig:actions}, allowing us to more efficiently illustrate wandering across the entire chaotic zone.}
\label{fig:jwander}
\end{figure}

Figure \ref{fig:jwander} displays the quasi-random wandering (``adiabatic chaos'', Neishtadt 1991) of the
action of a model system with $\alpha =0.2$, $\kappa=0.1$ during
approximately 5000 separatrix crossings. ``Sticking'' near the top and
bottom of the chaotic zone is a consequence of the smaller size of the
action jumps in these regions as predicted by Eq. (\ref{eq:tim});
$\dot A/A=0$ at the extrema of $A(t)$. 

\begin{figure}
\begin{center}
\includegraphics[width=8cm]{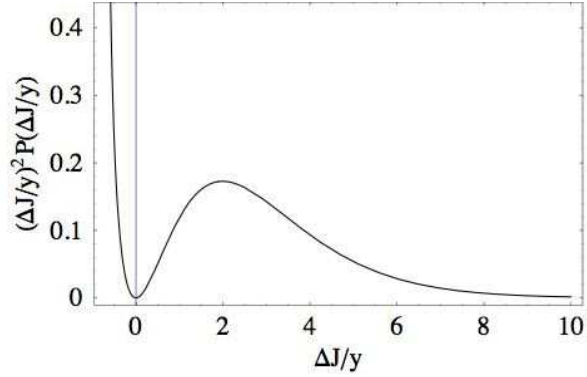}
\caption{The system action diffuses through the chaotic zone due to quasi-random jumps in action at separatrix crossings. The jump size is given by Eq. \ref{eq:tim}, and assuming a uniform distribution of $\sin (\psi_s/2)$, we can construct a probability distribution $P(\Delta J)$ for jump size. The contribution to the diffusivity from a given jump size $\Delta J$ is proportional to $(\Delta J)^2 P(\Delta J)$. This quantity is plotted in the figure, normalizing jump size to $y=4\dot A/A$. Jumps of more than 2 times this size contribute about half of the diffusivity while only contributing about 4 \% by number.}
\label{fig:djdist}
\end{center}
\end{figure}

Consider the distribution of jump sizes.  When crossing a separatrix,
the system trajectory passes over a potential peak. Smaller margins of
clearance correspond to slower passages and result in larger jumps in
the action. The term $\sin (\psi_s/2)$ is proportional to the difference
between the system's energy and the separatrix energy when the system
makes its last passage through a potential trough prior to crossing a
separatrix. We construct the probability distribution $P(\Delta J)$ of
jump sizes under the assumption that $\sin (\psi_s/2)$ is uniformly
distributed between $\pm 1$. Over many separatrix crossings the
action performs a quasi-random walk in which the diffusion constant is
proportional to $\int d(\Delta J) P(\Delta J)(\Delta J)^2$.  From the
plot of the integrand in Figure \ref{fig:djdist}, we see that jumps
larger than $8 \dot A/A$, which comprise about 4\% by number,
contribute about half of the diffusivity.

Integrating over jump size, we obtain a diffusion coefficient:
\begin{equation}
D \simeq 0.82 \left(\frac{4 \dot A}{A}\right)^2 \frac{\kappa}{\pi},
\label{eq:d}
\end{equation}
where $\pi/\kappa$ is the mean time between separatrix crossings.
Taking values appropriate to the center of the chaotic zone, we
estimate the zone crossing time
\begin{equation}
t_c \sim \frac{J_{\rm cz}^2}{D} \sim 15,000,
\label{eq:tc}
\end{equation}
in reasonable agreement with that observed in Figure
\ref{fig:jwander}.



The pendulum mostly librates or mostly circulates, according to
whether its action is close to the minimum or maximum value in the
chaotic zone. This is illustrated in Figure \ref{fig:hilo}. As
expected, the system samples each type of behavior.

\section{Prometheus and Pandora}
\label{sec:panprom}

\begin{figure}
\begin{center}
\includegraphics[width=0.5 \textwidth]{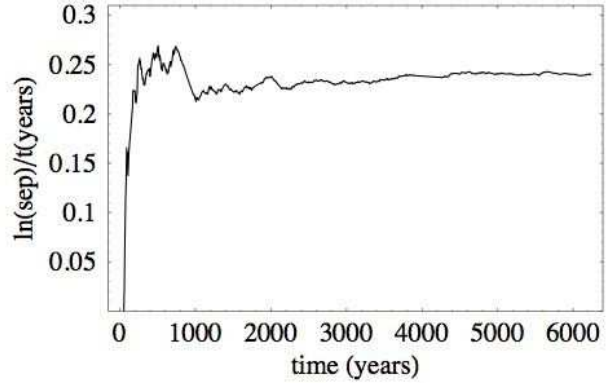}
\end{center}
\caption{Determination of the Lyapunov exponent for the
Prometheus-Pandora system: this plot is analogous to Fig. \ref{fig:model_sep}b for the model parametric pendulum, in which the Lyapunov exponent is given by the late-time value of $\ln({\rm separation})/{\rm time}$.}
\label{fig:pp_lyap}
\end{figure}

Now we return to the Prometheus-Pandora system and detail similarities
between its behavior and that of a parametric pendulum. Although our
interpretation of the satellites' motions rests on the pendulum
analogy, we integrate the full equation of motion (Eq. \ref{eq:full})
for $\psi(t)$ and the linear equation for the tangent vector
$\Delta\psi(t)$ derived from it. Because the system is chaotic, it
explores the entire accessible region of phase space for arbitrary
initial data. We find it most convenient to adopt the initial data
used by GR03b for easier comparison between their results and ours.  A
long integration yields a Lyapunov exponent $\ell \sim 0.3$ yr$^{-1}$,
in agreement with that found by GR and other authors (see Figure
\ref{fig:pp_lyap}).

\begin{figure}
\begin{center}
\includegraphics[width=0.75 \textwidth]{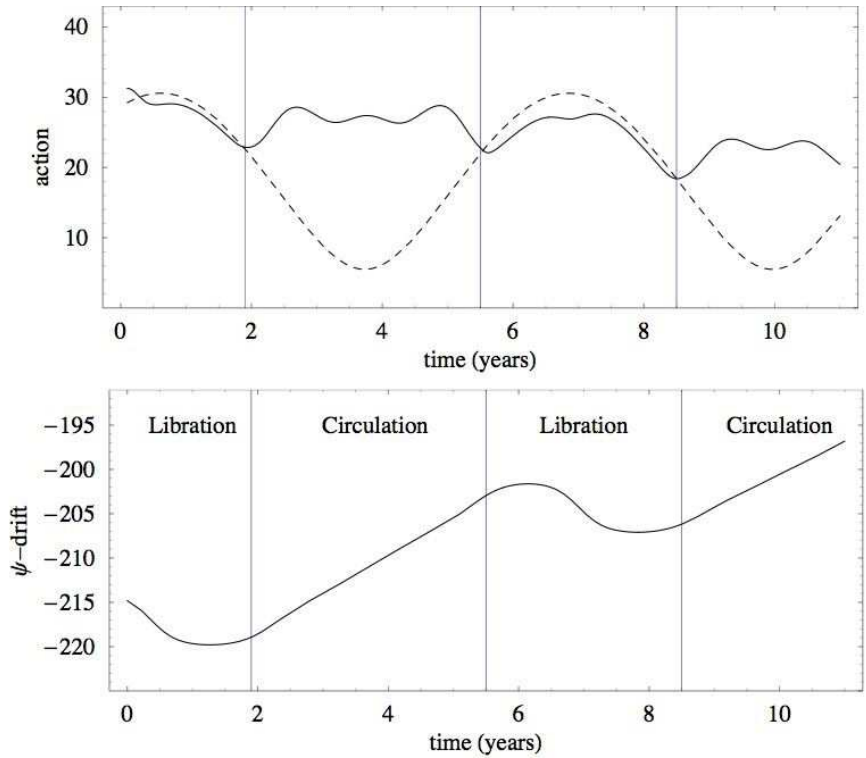}
\end{center}
\caption{Action
  and $\psi$ for the Prometheus-Pandora system for the first 12 years of
  integration, showing changes in the regime of motion at separatrix
  crossings, as indicated by vertical lines. The dashed line denotes the separatrix action. The unit of action for the Prometheus-Pandora system is rad$^2$ yr$^{-1}$ throughout.}
\label{fig:libcirc}
\end{figure}

The variations of $J$ and $\psi$ for the satellite system (plotted
in Figure \ref{fig:libcirc}) are reminiscent of those of the analogous
variables pertaining to the parametric pendulum as illustrated in
Figure \ref{fig:model_plc}. Similar patterns of switching between
libration and circulation are exhibited by each system. Fractional
variations of the action are larger in the satellite system because of
its lower adiabaticity.

\subsection{Where separatrix crossings occur}
\begin{figure}
\includegraphics[width=0.5 \textwidth]{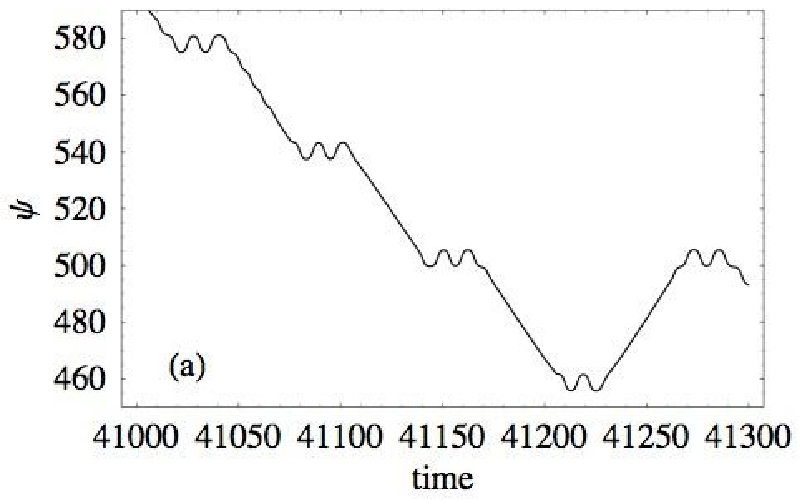}
\includegraphics[width=0.5 \textwidth]{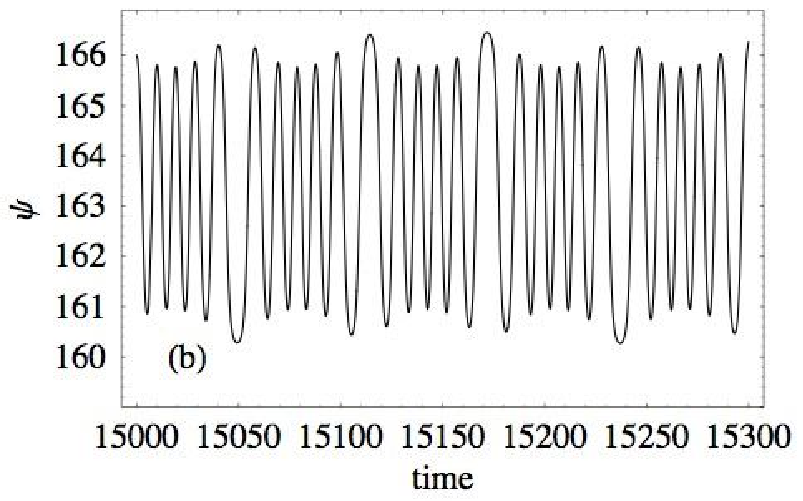}
\caption{Evolution of the pendulum in Fig. \ref{fig:jwander}, showing the behavior of $\psi$ when the action is close to the (a) top and (b) bottom of the chaotic zone.}
\label{fig:hilo}
\end{figure}

GR03ab and subsequent authors state that jumps in the mean motions
occur near apse antialignment. Apse antialignment corresponds to the
time of maximum amplitude of the potential. We see in Figure
\ref{fig:libcirc} that separatrix crossings currently occur in that
region. A similar behavior is seen in Figure \ref{fig:hilo}a, in which
the action is close to its maximum value in the chaotic zone. Jumps in
mean motion correspond to intervals spent in libration amidst
stretches of circulation during which $\psi$ increases monotonically.
Changes in the slope, $\dot \psi$, from one episode of circulation to
the next reflect changes in the action at the intervening separatrix
crossings.  The jumps are not \emph{due} to chaos; rather they are due
to the separatrix crossings, which \emph{also} give rise to
chaos. Changes in the circulation rate can be predicted, but only with
an error that increases at every separatrix crossing.

\begin{figure}
\begin{center}
\includegraphics[width=0.7 \textwidth]{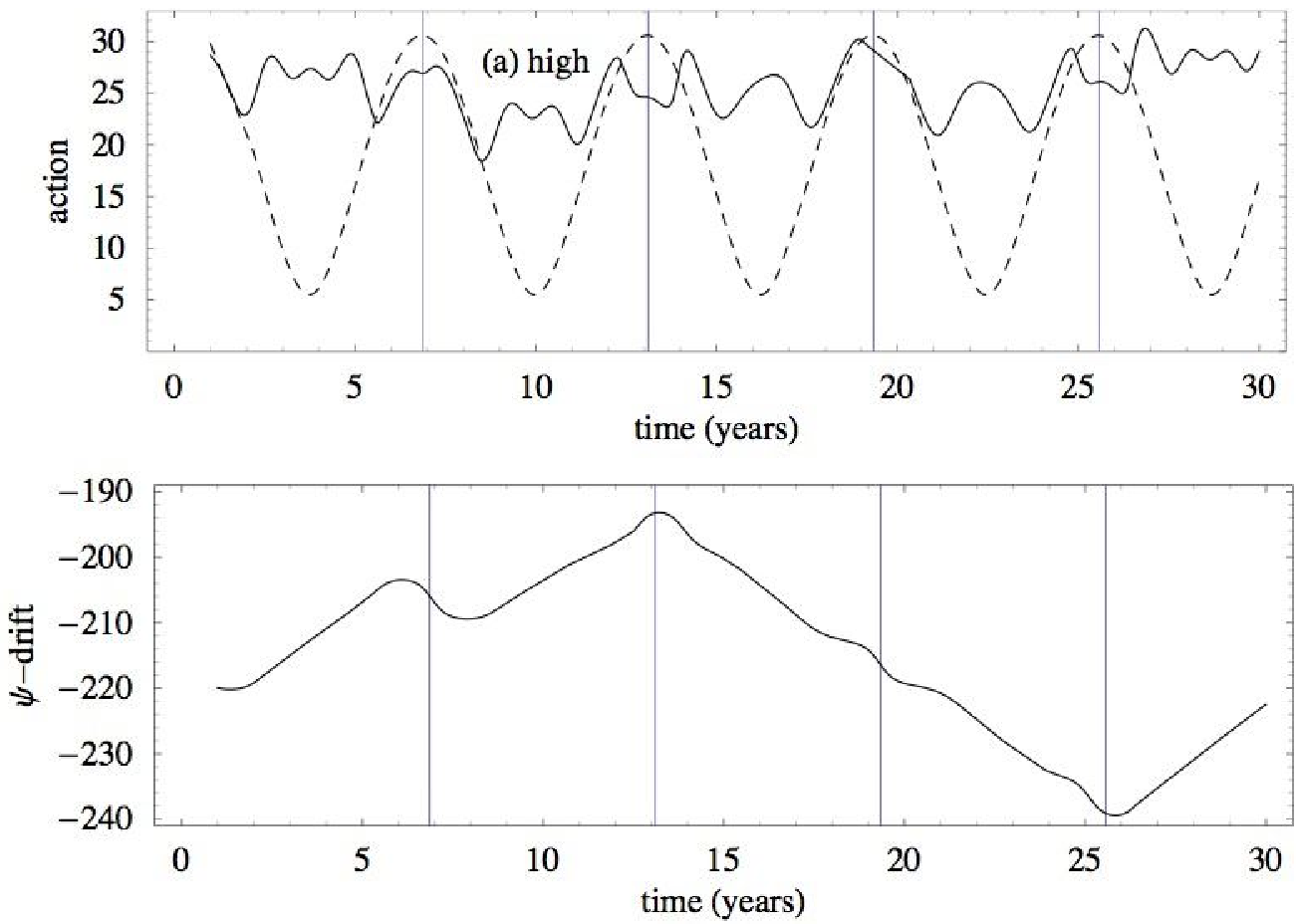}
\includegraphics[width=0.7 \textwidth]{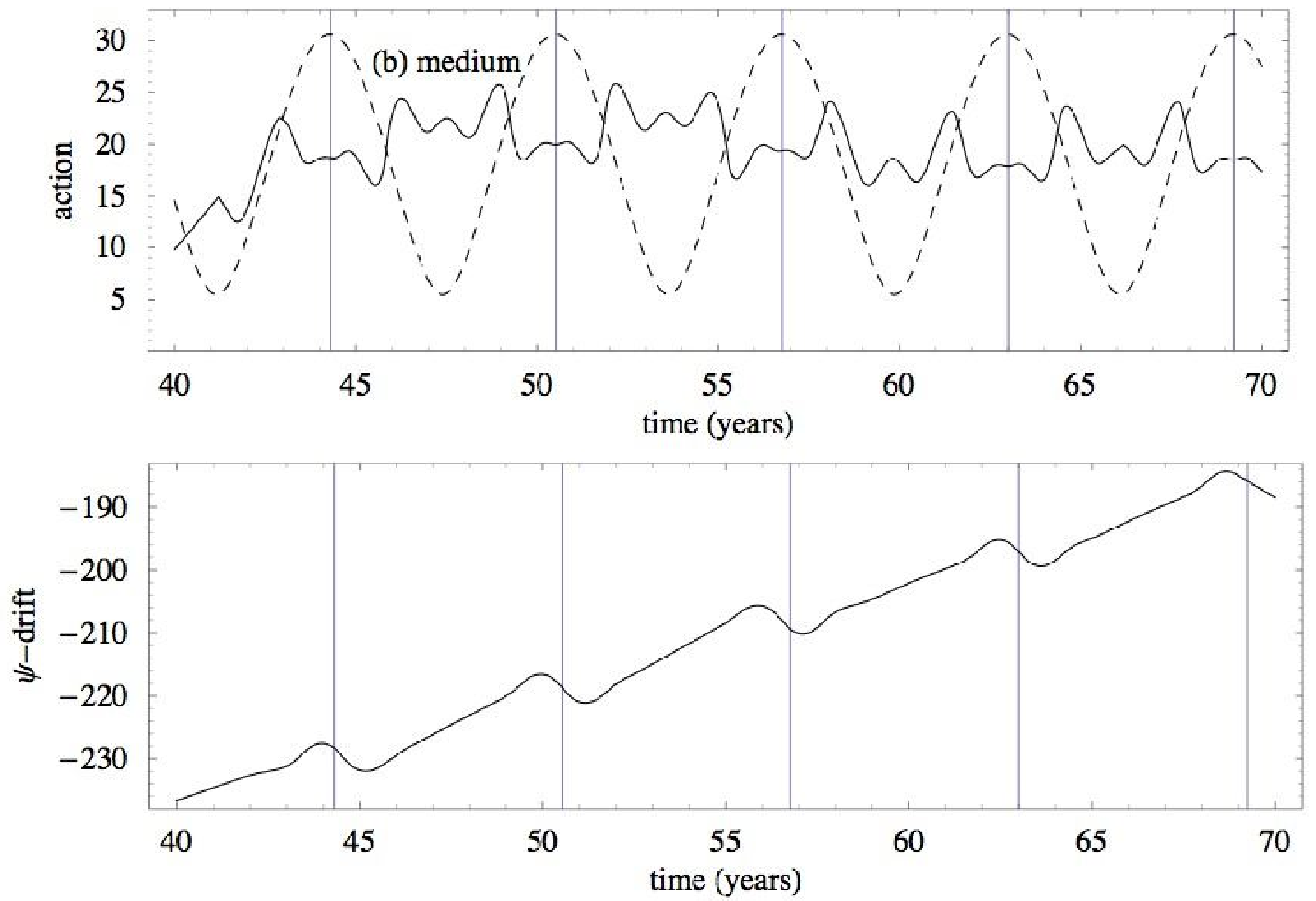}
\includegraphics[width=0.7 \textwidth]{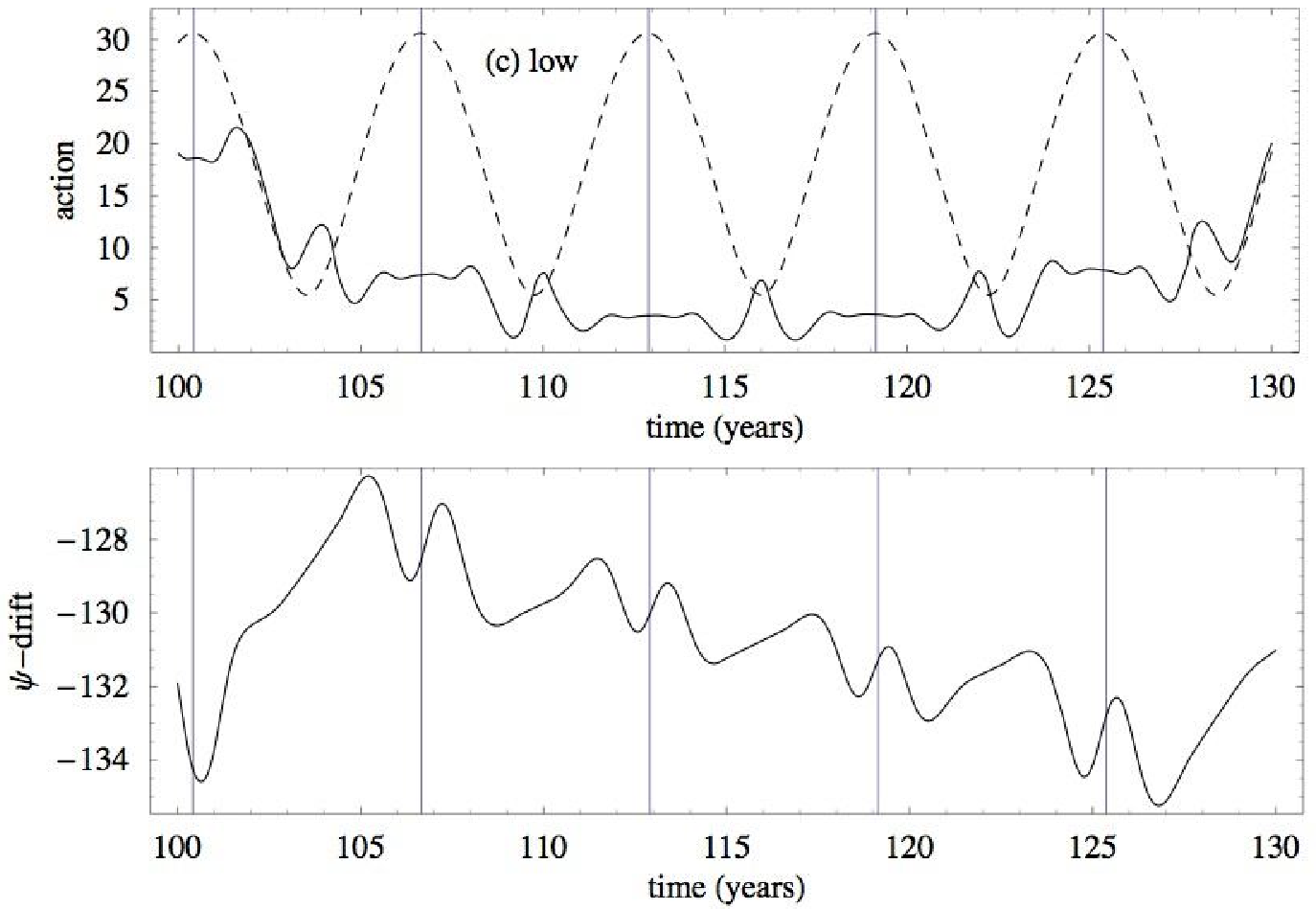}
\end{center}
\caption{Examples of high,
  medium, and low action states for the Prometheus-Pandora system. Vertical lines indicate
  times of apse antialignment, \emph{not} separatrix crossings, which themselves occur
  at different precessional phases in these three cases. Dashed lines are separatrix action.}
\label{fig:range}
\end{figure}

A chaotic system explores the entire chaotic zone in phase space.  We
expect the Prometheus-Pandora system to sample states that are unlike
its current state in which the separatrix crossings occur near apse
antialignment as seen in Figure \ref{fig:hilo}a. Averaged over time,
separatrix crossings should not be restricted to a narrow range of
precessional phase. In particular, states in which the separatrix
crossings take place near apse alignment, such as shown in Figure
\ref{fig:hilo}b, must also be sampled. Results from our integrations
displayed in Figure \ref{fig:range} confirm this expectation; the
system experiences both states in which it is mostly in circulation and
those in which it is mostly in libration.  Separatrix crossings, accompanied by sudden
changes in the behavior of $\psi(t)$, happen at all precessional
cycles. Switching between regimes of behavior does not require a ``kick'' as such; the requirement is that the potential change so that the system action moves to the other side of the separatrix action. This can happen as easily near apse alignment as near apse antialignment.

\begin{figure}
\includegraphics[width=\textwidth]{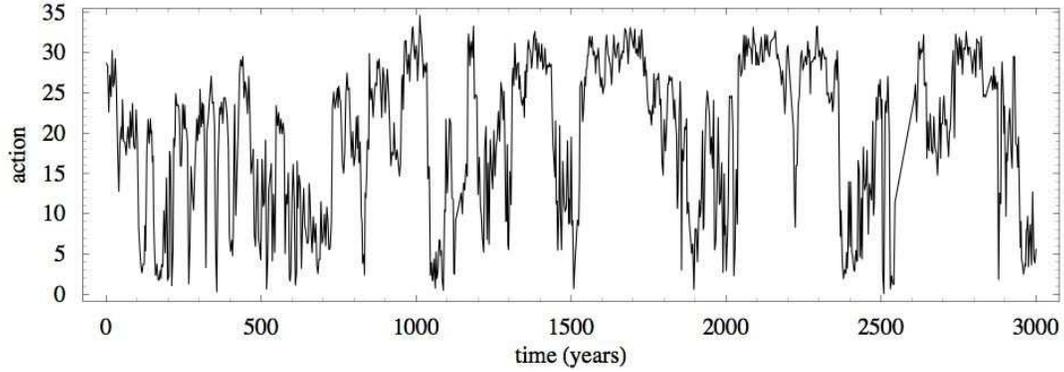}
\caption{Wandering of the action of the Prometheus-Pandora system: analogous to Fig. \ref{fig:jwander} for a model pendulum, this clearly delineates the chaotic zone.}
\label{fig:pp_jwander}
\end{figure}

A long integration confirms that the action wanders throughout the
chaotic zone (Figure \ref{fig:pp_jwander}, cf. also Figure
\ref{fig:jwander}). Eqs. (\ref{eq:d}) and (\ref{eq:tc}) predict that it takes $\sim$ 170
years for the action to diffuse all the way across the zone. Due to the weak adiabaticity in this case, a single large jump
can span a large fraction of the total range of action. Flipping between
moderately high and moderately low action states can therefore occur in a much shorter time,
as seen in Figure \ref{fig:pp_jwander}.

\begin{figure}
\begin{center}
\includegraphics[width=4in]{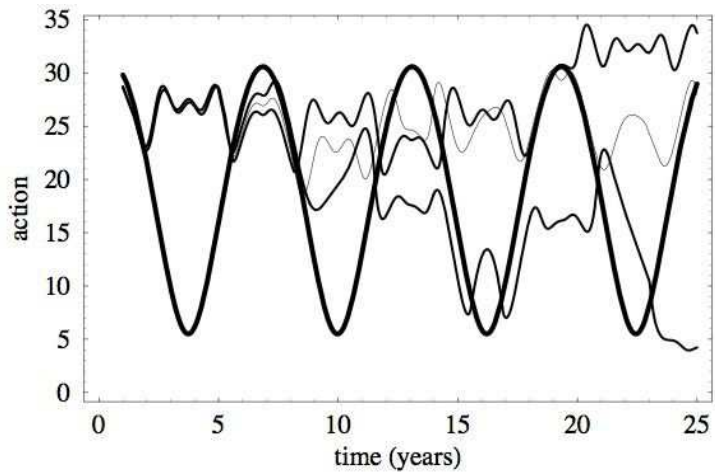}
\caption{We start with three sets of initial conditions of Prometheus and Pandora corresponding to 0.2$^{\circ}$ uncertainty in the longitude of each satellite, and plot the action (along with the separatrix action, thick line) for each of these trajectories. In as little as 15 years, one of these trajectories has transitioned from a high to a low action state.}
\label{fig:whenchange}
\end{center}
\end{figure}

Within measurement errors, a variety of orbits fit the best orbital
data for Prometheus and Pandora.  Renner et al. (2005) find that after
two separatrix crossings (their 2004 point), there is an uncertainty
of 0.2$^\circ$ in the longitude of each satellite.\footnote{This is
about the same as the size of a single \emph{HST} error box at
Saturn.}  Starting from this level of positional uncertainty and
holding all other orbital elements fixed, we integrate forward in time
a set of 3
trajectories spanning the error range. The results are
shown in Figure \ref{fig:whenchange}, in which the action for each
trajectory is plotted against the separatrix action. In only 15 years
from the start of our integration, one of the trajectories has already
transitioned from a state of high to low action (due to experiencing a
large jump). This illustrates that
only a small uncertainty in positions can rapidly amplify to give a
qualitative change in system behavior.


\subsection{Limitations of the analogy}
\label{sec:probs}

The Prometheus-Pandora system is not strongly adiabatic, since only
about 3 periods of oscillation take place during the modulation
of the potential. Because the amplitude of the potential varies
substantially over the period of motion, we expect our frozen
potential assumption to lead to uncertainties of order half a period
of the motion in the positioning of the separatrix crossings. However,
we have seen that the crossing positions are adequately located by
this method.

A more serious problem is the fact that we do not fulfill the
condition $\ddot \phi \ll |V(t)|$ throughout all of the precessional
cycle. The dashed line plotted in Figure \ref{fig:pot} shows the
magnitude of the $\ddot \phi$ term in comparison with
$|V(t)|=|\partial V/\partial \psi|$. For about one third of the time
(around apse alignment), we do not fulfill this condition, and
the pendulum equation does not describe the system well at all. During
these times, the system is essentially governed by $\ddot \Psi \simeq
-\ddot \phi$, so $\Psi$ appears to circulate. Perhaps then we cannot
define a separatrix crossing near apse alignment. We do find some
evidence (not detailed in this paper) that greater jumps in the
separation of trajectories occur near apse antialignment than near
apse alignment. Undeniably however, changes in the evolution of
$\psi(t)$ occur throughout the precessional cycle (Figure
\ref{fig:range}), and this statement is independent of the pendulum
analogy, since it is based on integrations of the full equation of
motion, Eq. (\ref{eq:full}). The behavior of the orbits between apse
antialignments certainly cannot always be fitted using freely
precessing ellipses.



\section{Summary and conclusions}
\label{sec:conc}
The equations of motion of the Prometheus-Pandora resonant system are
well represented by those of a parametric pendulum. Despite the low
degree of adiabaticity of the system, the same regimes of behavior
exist as found in a more adiabatic model system. Using this analogy,
we have explained the nature of the ``kinks'' in mean longitude seen in
both observations and simulations as short episodes of libration
between long stretches of circulation. We caution that the system will
not continue indefinitely to display periods of drift with intervening
kinks: in as little as 15 years, the evolution of the mean motions
could be drastically different. Sudden changes in behavior (and
increased uncertainties in predictions) can happen far from times of
apse antialignment.

\section{Acknowledgments}

This research was supported by NASA grant PGG 344-30-55-07 and NSF
grant AST 00-98301. AJF thanks the Institute for Advanced Study for
its hospitality. 

\vfill\eject

\newpage
Tables:

\begin{table}
\begin{center}
\begin{tabular}{crr}
\hline
Quantity & Prometheus & Pandora\\
\hline
Symbols & unprimed & primed\\
m/M & $5.80 \times 10^{-10}$& $3.43 \times 10^{-10}$\\
$\lambda$ ($^\circ$) & 188.53815 & 82.14727 \\
$n$ ($^\circ$ s$^{-1}$) & $6.797331 \times 10^{-3}$ & $6.629506 \times
10^{-3}$\\
$e$ & $2.29 \times 10^{-3}$&$4.37 \times 10^{-3}$\\
$\varpi$ ($^\circ$)&212.85385&68.22910\\
$\dot \varpi$ ($^\circ$ s$^{-1}$) & $3.1911 \times 10^{-5}$ & $3.0082
\times 10^{-5}$\\
\hline
\end{tabular}
\caption[The orbits and properties of Prometheus and
  Pandora]{Properties of Prometheus and Pandora and of their orbits,
  from GR03b}
\label{tab:eqterms}
\end{center}
\end{table}
\newpage
Tables (continued):

\begin{table}
\begin{center}
\begin{tabular}{ccr}
\hline
Resonance $q$ & Period (yr) $=-2\pi/\dot\psi_q$ & Coefficient $C_q$\\
\hline
1 & 1.078&$-1.08\times 10^{-3}$\\
2 & 1.303&$6.26\times 10^{-3}$\\
3 & 1.648&$-1.21\times 10^{-2}$\\
4 & 2.239&$7.82\times 10^{-3}$\\
\hline
\end{tabular}
\caption[Components of the 121:118 mean motion resonance]{Properties
  of the four components of the 121:118 mean motion resonance, from
  GR03b}
\label{tab:res}
\end{center}
\end{table}

\newpage
\listoffigures


\end{document}